\documentclass[twocolumn,showpacs,preprintnumbers,amsmath,amssymb]{revtex4-1}
\usepackage{graphicx}
\usepackage{rotating}
\begin{document}

\renewcommand{\theequation}{\thesection.\arabic{equation}}

\newcommand{\re}{\mathop{\mathrm{Re}}}

\newcommand{\be}{\begin{equation}}
\newcommand{\ee}{\end{equation}}
\newcommand{\bea}{\begin{eqnarray}}
\newcommand{\eea}{\end{eqnarray}}

\title{Regularizing cosmological singularities by varying physical constants.}

\author{Mariusz P. D\c{a}browski}
\email{mpdabfz@wmf.univ.szczecin.pl}
\affiliation{\it Institute of Physics, University of Szczecin, Wielkopolska 15, 70-451 Szczecin, Poland}
\affiliation{\it Copernicus Center for Interdisciplinary Studies,
S{\l }awkowska 17, 31-016 Krak\'ow, Poland}

\author{Konrad Marosek}
\email{k.marosek@wmf.univ.szczecin.pl}
\affiliation{\it Institute of Physics, University of Szczecin, Wielkopolska 15, 70-451 Szczecin, Poland}
\affiliation{\it Chair of Physics, Maritime University, Wa{\l}y Chrobrego 1-2, 70-500 Szczecin, Poland}

\date{\today}

\input epsf

\begin{abstract}
Varying physical constant cosmologies were claimed to solve standard cosmological problems such as the horizon, the flatness and the $\Lambda$-problem. In this paper, we suggest yet another possible application of these theories: solving the singularity problem. By specifying some examples we show that various cosmological singularities may be regularized provided the physical constants evolve in time in an appropriate way.

\end{abstract}


\maketitle

\section{Introduction}
\label{intro}
\setcounter{equation}{0}

One of the most intriguing problems in cosmology is the problem of singularities. They are very well-defined in relativity and are shown to appear under quite general conditions of geodesic incompletness and a blow-up of various geometrical and physical quantities \cite{HE}. Up to a very recently, the main concern of cosmologists was the big-bang singularity in the past which seemed to be unavoidable both in relativity and also in extended scalar field theories \cite{BV}. A lot of generalized theories which included gravity were proposed in order to avoid big-bang. Among them the superstring and the brane theories \cite{string,brane}, loop quantum gravity \cite{LQC}, higher-order gravity \cite{f(R)}, and many others. The main achievement of such approaches was the extension of the evolution of the universe through a big-bang singularity like in the pre-big-bang \cite{PBB} and the cyclic \cite{cyclic} scenarios.

After the discovery of the accelerated expansion of the universe \cite{supernovaeold,supernovaenew,supernovaesupernew}, deeper studies of the phenomenon of the dark energy showed the plethora of new singularities (``exotic'' singularities) different from big-bang. Firstly, a big-rip associated with the phantom dark energy was investigated \cite{phantom}, and further classified as type I \cite{nojiri}. Then, a sudden future singularity (SFS or type II) was proposed \cite{SFS} as well as numerous
other types such as: generalized sudden future singularities (GSFS), finite scale factor singularities (FSF or type III),  big-separation singularities (BS or type IV) and $w$-singularities (type V) \cite{wsing,type0V}. The singularities which fall outside this classification (with perhaps a big-bang as type 0 \cite{type0V}) are curvature singularities with respect to a parallelly propagated basis (p.p. curvature singularites) which show up as directional singularities \cite{LFJ2007} and also intensively studied recently: the little-rip singularities \cite{LRip}, and the pseudo-rip singularities \cite{PRip}.  All the above singularities are characterized by violation of all, some or none of the energy conditions which results in a blow-up of all or some of the appropriate physical quantities such as: the scale factor, the energy density, the pressure, and the barotropic index (for a review see Ref. \cite{aps10}). In order to be clear, we remind that there are three energy conditions: the null ($\varrho c^2 + p \geq 0$), weak ($\varrho c^2 \geq 0$ and $\varrho c^2 + p \geq 0$), strong ($\varrho c^2 + p \geq 0$ and $\varrho c^2 + 3p \geq 0$), and dominant energy ($\varrho c^2 \geq 0$, $-\varrho c^2 \leq p \leq \varrho c^2$) (here $c$ is the speed of light, $\varrho$ - the mass density in kg m$^{-3}$, and $p$ - the pressure). One can also define $\varepsilon \equiv \varrho c^2$ as the energy density which has the same unit as pressure, i.e., $J m^{-3} = N m^{-2} = kg m^{-1} s^{-2}$.

It emerged fascinating that some of these singularities are weaker than big-bang (for example particles \cite{lazkoz} and even extended objects \cite{adam} may pass through them) and may appear in the very near future \cite{SFSobserv,FSFobserv}. Then, it is interesting to discuss if there are any physical reasons which can ``weaken'' or just remove a big-bang (or other - ``stronger'') singularity from the evolution of the universe. Our suggestion here is to make use of the time-evolution of the physical constants combined with the dynamical evolution of particular models of the universe. It has been recently shown \cite{houndjo} that the quantum effects may change the nature of the sudden future singularity as well as the big-rip and the finite scale factor singularity. It is then reasonable to think of the time-varying physical constants to do the job as well.

The idea of variation of physical constants has been established widely in physics both theoretically and experimentally. From the theoretical side the early ideas of Weyl \cite{weyl} and Eddington \cite{eddington} were most successfully followed by Dirac's Large Number Hypothesis \cite{dirac} from which it was concluded that the gravitational constant should change in time as $G(t) \propto 1/t$. This led to the scalar-tensor gravity theory developed by Brans and Dicke \cite{bd} who followed the ideas of Jordan \cite{jordan}. These ideas were further embedded into superstring theories in which the coupling constant of gravity became running during the evolution of the early universe \cite{string}. In fact, a lot of physical constants such as the gravitational constant $G$, the charge of the electron $e$, the velocity of light $c$, the proton to electron mass ratio $m_p/m_e$, and the fine structure constant $\alpha$ are interrelated \cite{uzan,barrowbook} and the variation of one of them may be associated with variation of others. However, apart from Brans-Dicke type of gravitational constant evolution models, the most popular theories which admit physical constants variation are the varying speed of light theories \cite{VSL} and varying alpha (fine structure $\alpha$) theories \cite{alpha}. It has been shown that both of these theories allow the solution of the standard cosmological problems such as the horizon problem, flatness problem, and the $\Lambda-$problem. Here, we will apply these theories to solve yet another problem - the singularity problem.

The paper is organized as follows. In Section \ref{models} we discuss the basics of varying constants models using the new form of the scale factor which encompasses quite a few types of singularities after a specific choice of its parametrization. In Section \ref{regular} we show the examples of regularization the singularities due to the time-variation of physical constants. In Section \ref{conclusion} we give our conclusions.

\section{Varying constants models}
\setcounter{equation}{0}
\label{models}

Following the Refs. \cite{VSL}, we consider the Friedmann universes within the framework of varying speed of light theories (VSL) and varying gravitational constant theories. The field equations read as
\bea \label{rho} \varrho(t) &=& \frac{3}{8\pi G(t)}
\left(\frac{\dot{a}^2}{a^2} + \frac{kc^2(t)}{a^2}
\right)~,\\
\label{p} p(t) &=& - \frac{c^2(t)}{8\pi G(t)} \left(2 \frac{\ddot{a}}{a} + \frac{\dot{a}^2}{a^2} + \frac{kc^2(t)}{a^2} \right)~,
\eea
and the energy-momentum conservation law is
\be
\label{conser}
\dot{\varrho}(t) + 3 \frac{\dot{a}}{a} \left(\varrho(t) + \frac{p(t)}{c^2(t)} \right) = - \varrho(t) \frac{\dot{G}(t)}{G(t)}
+ 3 \frac{kc(t)\dot{c}(t)}{4\pi Ga^2}~.
\ee
Here $a \equiv a(t)$ is the scale factor, the dot means the derivative with respect to time $t$, $G=G(t)$ is time-varying gravitational constant, $c=c(t)$ is time-varying speed of light, and the curvature index $k=0, \pm 1$. In Ref. \cite{VSL} the barotropic equation of state for matter was assumed. Since we want to discuss more general cases in order to obtain various types of singularities, we will not assume any link between the energy density and pressure (i.e. the equation of state $p = p(\varepsilon)$.

In contrast to many references dealing with sudden future singularities \cite{SFS,SFSobserv}, which consider the scale factor
\be
\label{oldscalef}
a(t) = a_s \left[\delta + \left(1 - \delta \right) \left( \frac{t}{t_s} \right)^m -
\delta \left( 1 - \frac{t}{t_s} \right)^n \right]~,
\ee
with $\delta, t_s, a_s, m, n$ being constants, we propose a new form of the scale factor, which after appropriate choice of parameters admits big-bang, big-rip, sudden future, finite scale factor and $w$-singularities and reads as
\be \label{newscalef} a(t) = a_s \left( \frac{t}{t_s} \right)^m \exp{\left( 1 - \frac{t}{t_s} \right)^n} ~, \ee
with the constants $t_s, a_s, m, n$.
Its expansion around $t \approx t_s$ gives (cf. \cite{SFSobserv})
\be
a(t) = a_s - m a_s \left( 1 - \frac{t}{t_s} \right)~.
\ee
Notice that in equation (\ref{newscalef}), $a_s$ has the unit of length and all the terms are dimensionless. The scale factor is zero $(a=0)$ at $t=0$ (a big-bang singularity), and it is a constant $a=a_s$ at $t=t_s$. The first and second derivatives of the scale factor (\ref{newscalef}) are
\bea
\label{dotnew}
\dot{a}(t) &=& a(t) \left[ \frac{m}{t} - \frac{n}{t_s} \left( 1 - \frac{t}{t_s} \right)^{n-1} \right],~\\
\ddot{a}(t) &=& \dot{a}(t) \left[ \frac{m}{t} - \frac{n}{t_s} \left( 1 - \frac{t}{t_s} \right)^{n-1} \right] \nonumber \\
\label{ddotnew}
&+& a(t) \left[ -\frac{m}{t^2} + \frac{n(n-1)}{t^2_s} \left( 1 - \frac{t}{t_s} \right)^{n-2} \right].~
\eea
From (\ref{dotnew})-(\ref{ddotnew}), one can see that for $1 < n < 2$ $\dot{a}(0) = \infty$ and $\dot{a}(t_s) = ma_s/t_s =$const., while $a(t_s) = a_s$, $\ddot{a}(0) = \infty$ and $\ddot{a}(t_s) = - \infty$ ($p \to \infty$) and we have a sudden future singularity \cite{SFS}.


\begin{figure*}
    \begin{tabular}{ccc}
      \resizebox{60mm}{!}{\rotatebox{0}{\includegraphics{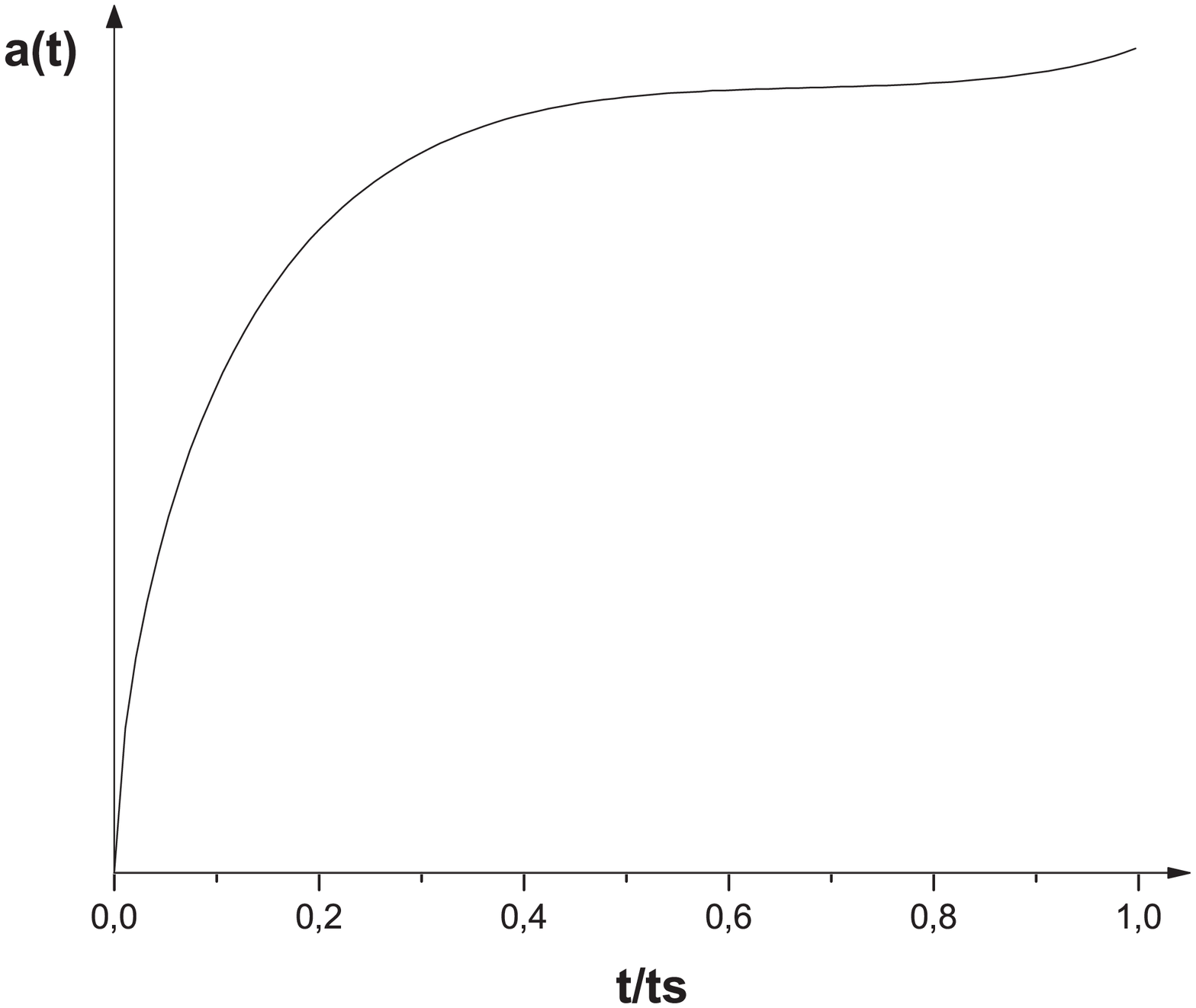}}} &
      \resizebox{60mm}{!}{\rotatebox{0}{\includegraphics{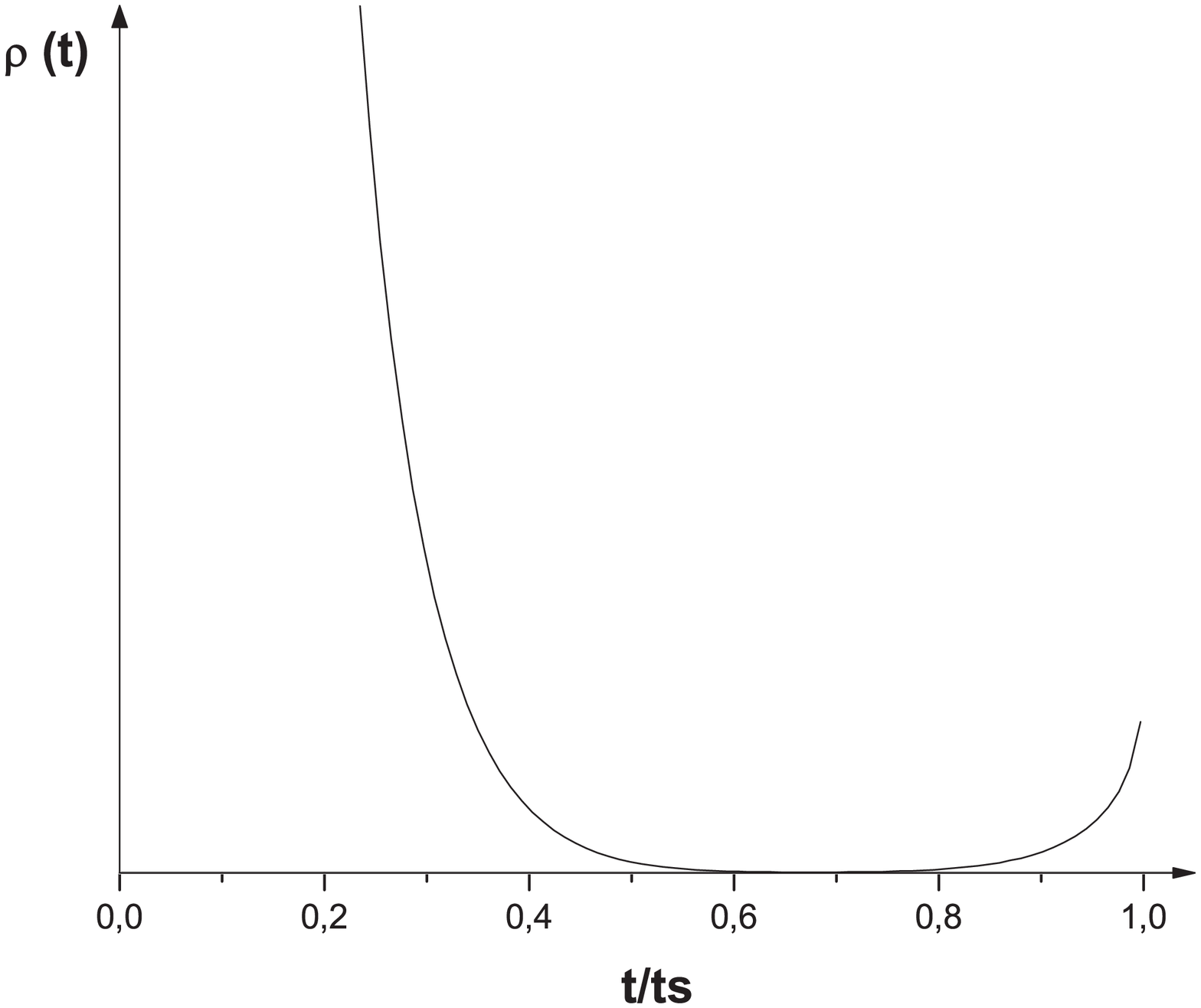}}}&
      \resizebox{60mm}{!}{\rotatebox{0}{\includegraphics{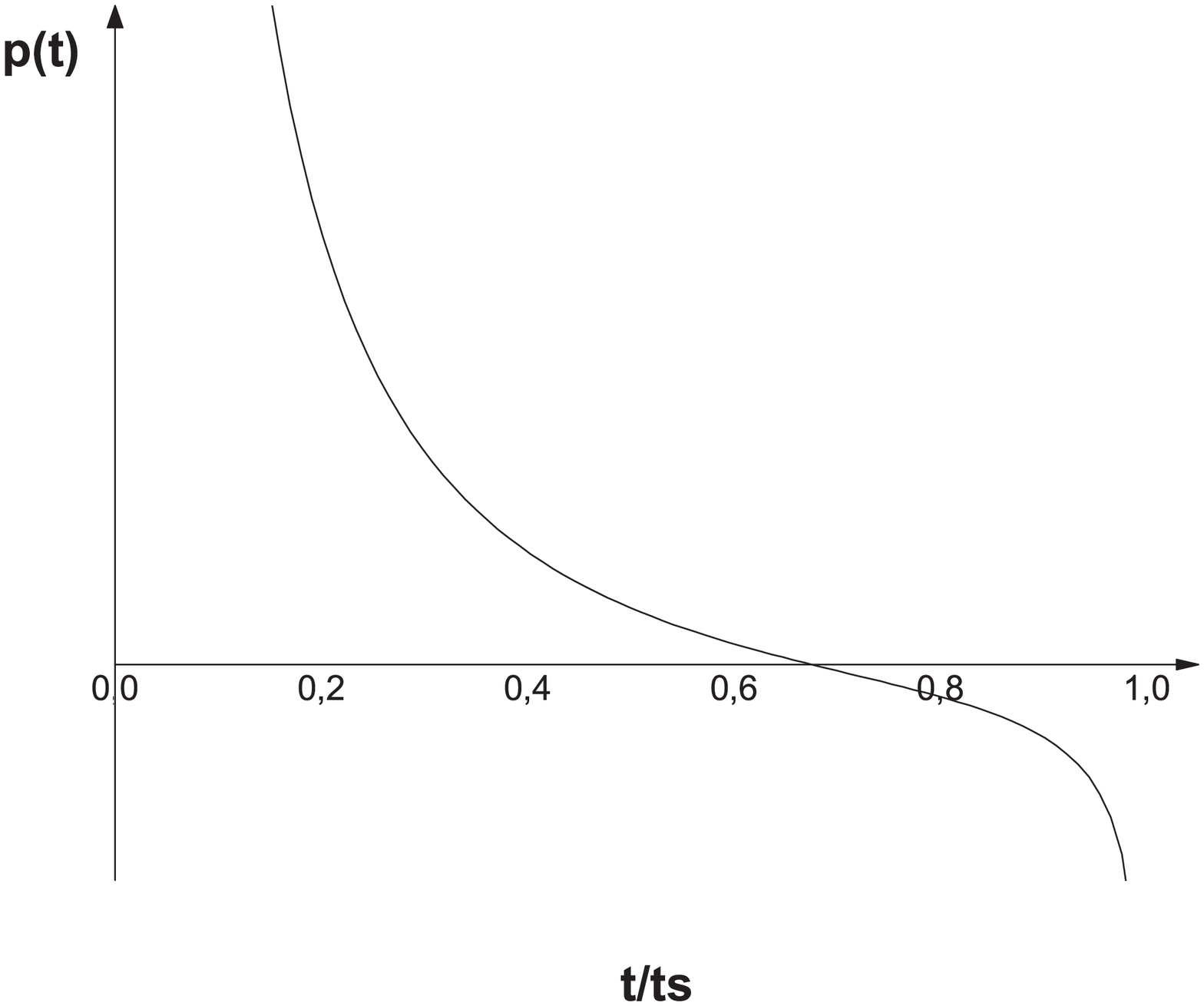}}}
       \\
      \resizebox{60mm}{!}{\rotatebox{0}{\includegraphics{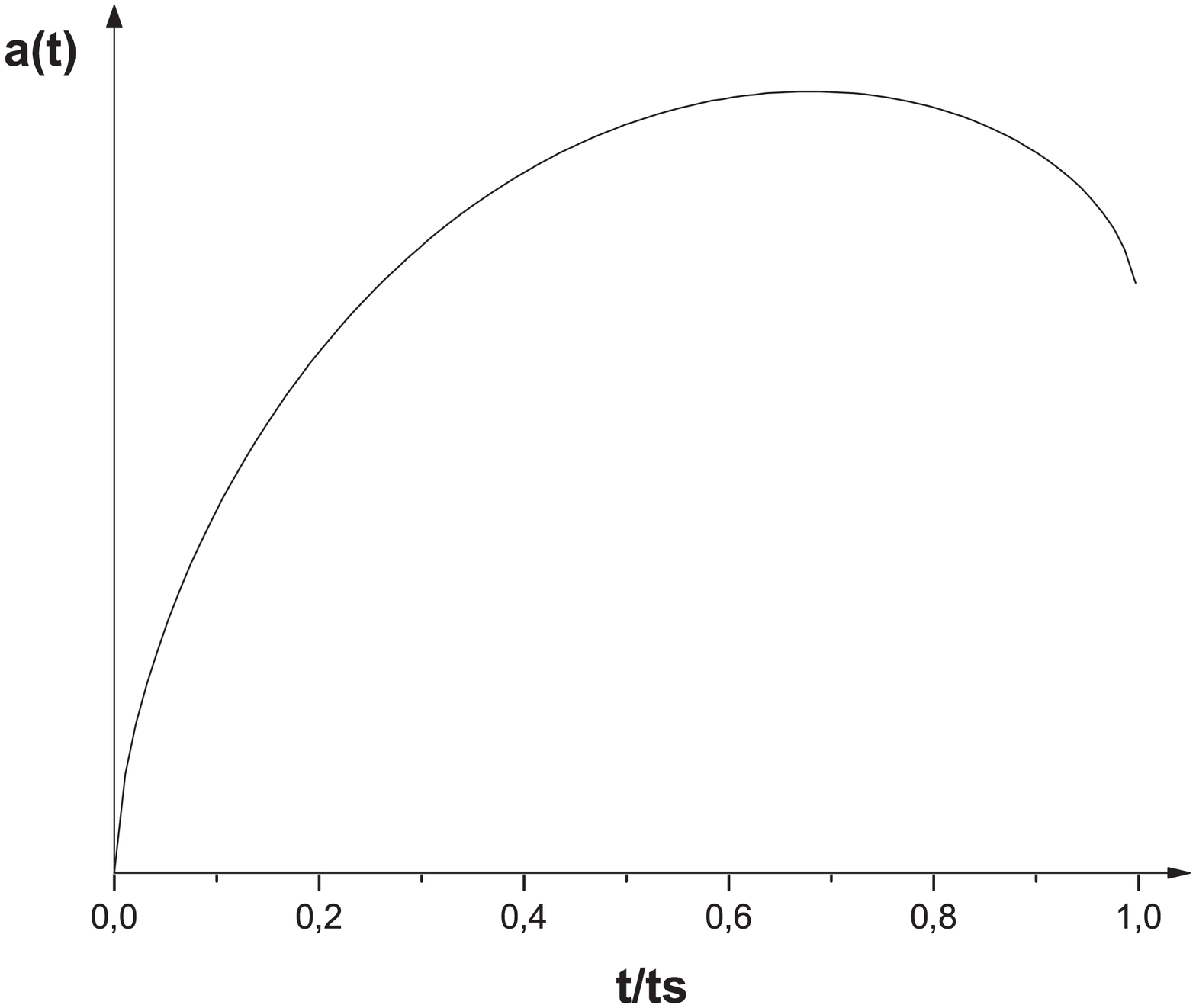}}} &
      \resizebox{60mm}{!}{\rotatebox{0}{\includegraphics{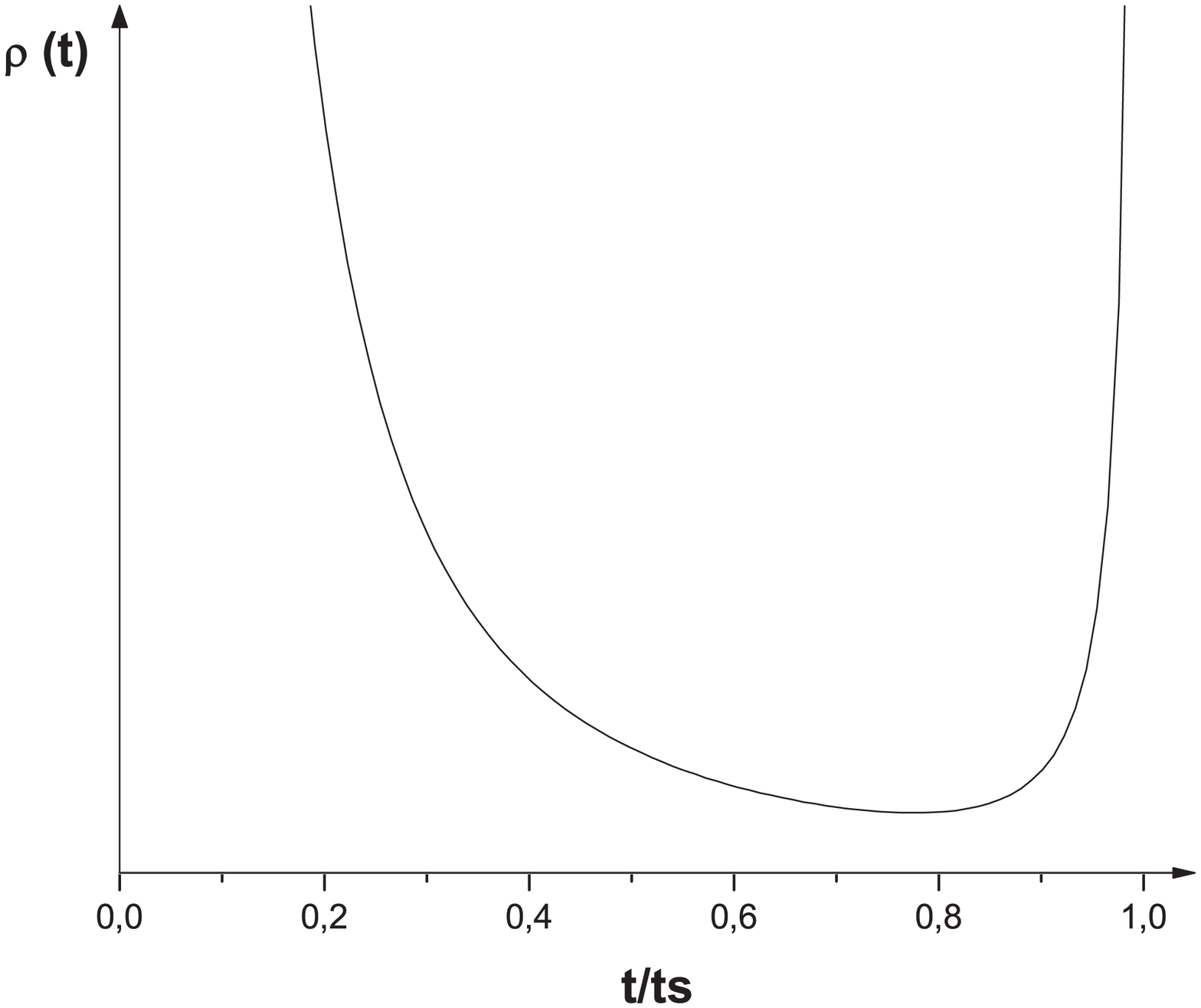}}} &
      \resizebox{60mm}{!}{\rotatebox{0}{\includegraphics{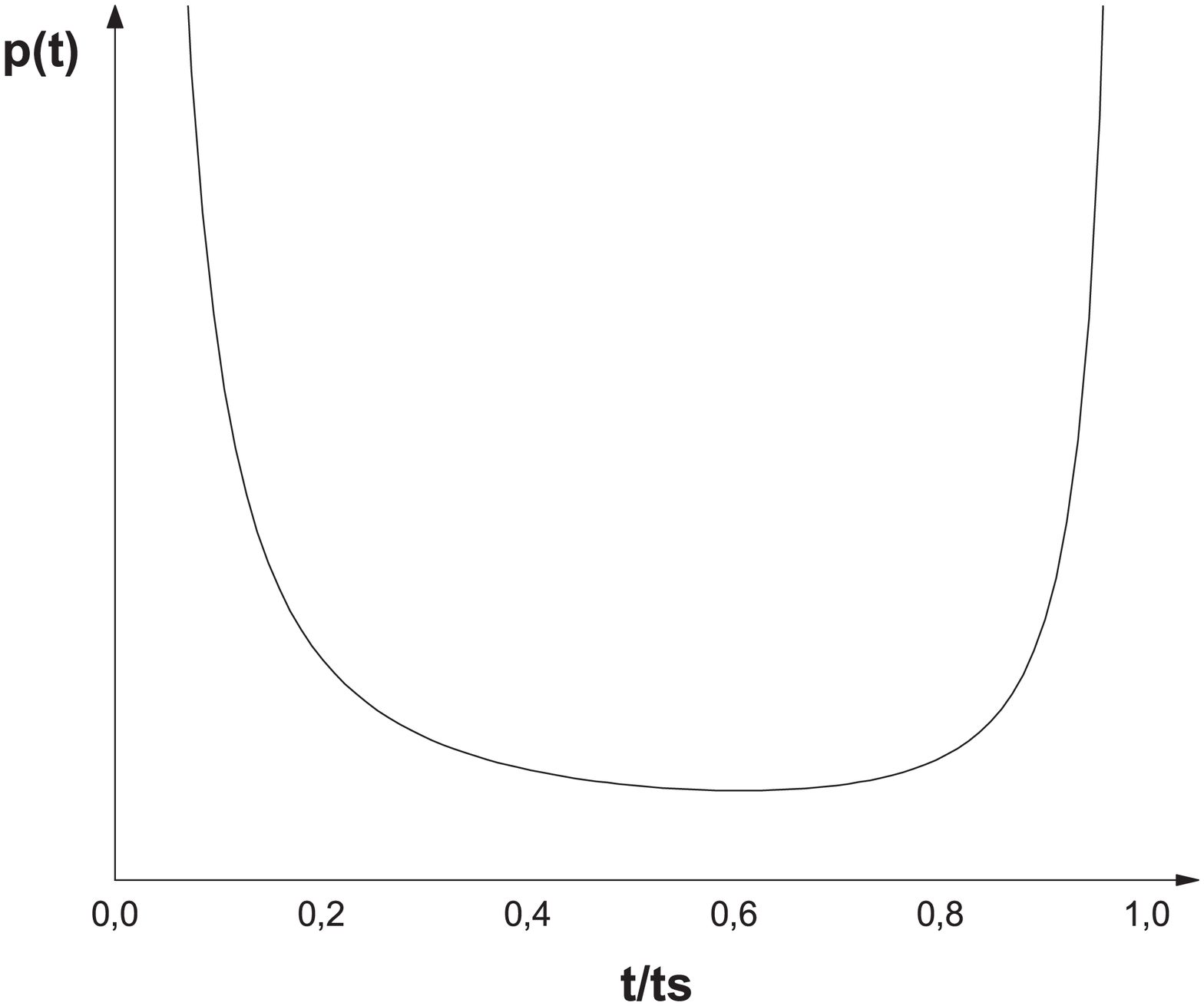}}}
       \\

    \end{tabular}

\caption{The plots of the scale factor $a(t)$, the energy density $\varrho(t)$, and the pressure $p(t)$ for the two specific models given by the scale factor (\ref{newscalef}). The first model is for the parameters $m=0.6$, $n=1.5$ and describes the sudden future singularity (SFS) scenario. The second model is for the parameters $m=0.6$ and $n=0.5$ and describes the finite scale factor singularity (FSF) scenario. }\label{fig1}
\end{figure*}

For a flat ($k=0$) Friedmann model it is possible to write down an explicit relation between the pressure and the energy density, though with a time-dependent barotropic index, in the form
\be
p_s(t) = w_s(t) \varepsilon_s(t) = w_s(t) \varrho(t) c^2(t)~~,
\label{pete}
\ee
where
\be
w_s(t) = \frac{1}{3} \left[2q(t) - 1 \right]~~,
\label{wute}
\ee
and $q(t) = - \ddot{a}a/\dot{a}^2$ is the (dimensionless) deceleration parameter.

Using (\ref{rho}), (\ref{p}), (\ref{dotnew}), and (\ref{ddotnew}) for the curvature index $k=0$ we have
\bea
\label{rhot}
\varrho(t) &=& \frac{3}{8\pi G(t)} \left[\frac{m}{t} - \frac{n}{t_s} \left( 1 - \frac{t}{t_s} \right)^{n-1} \right]^2~,\\
\label{pt}
p(t) &=& - \frac{c^2(t)}{8 \pi G(t)} \left[ \frac{m(3m-2)}{t^2} - 6 \frac{mn}{tt_s} \left( 1 - \frac{t}{t_s} \right)^{n-1} \right. \\
&+& \left. 3 \frac{n^2}{t^2_s} \left( 1 - \frac{t}{t_s} \right)^{2(n-1)} + 2 \frac{n(n-1)}{t^2_s} \left( 1 - \frac{t}{t_s} \right)^{n-2} \right]~. \nonumber
\eea

From (\ref{newscalef}), (\ref{rhot}) and (\ref{pt}) we can conclude that for $0 < m < 2/3$ we deal with a big-bang singularity and $a \to 0$, $\varrho \to \infty$, $p \to \infty$ at $t \to 0$, while for $m < 0$ we have a big-rip singularity with $a \to \infty$, $\varrho \to \infty$, $p \to \infty$ at $t=0$. Besides, it is clear that a sudden future singularity (SFS) which appears for $1<n<2$ at $t=t_s$ ($a=a_s$, $\varrho=$ const., $p \to \infty$) and a stronger \cite{tipler+krolak,lazkoz,LFJ2007} finite scale factor singularity (FSF) appears for $0<n<1$ at $t=t_s$ ($a=a_s$, $\varrho \to \infty$, $p \to \infty$). The example plots of these models are given in Fig. \ref{fig1}. In fact, for the former only the last term in the pressure of the type $(1-t/t_s)^{n-2}$ blows-up, while for the latter two more terms $(1-t/t_s)^{n-1}$ and $(1-t/t_s)^{2(n-1)}$ in (\ref{pt}) do \cite{tipler+krolak}. Such a choice also leads to a blow-up of the energy density (\ref{rhot}).

For the sake of further discussion, we will split the scale factor (\ref{newscalef}) into the two factors, one giving a big-bang singularity $a_{BB}$, and another giving an exotic singularity $a_{ex}$, as follows:
\be
\label{asplit}
a(t) = a_{BB} \cdot a_{ex}~,
\ee
where
\bea
\label{aBB}
a_{BB} &=& \left(\frac{t}{t_s} \right)^m~,\\
\label{aex}
a_{ex} &=& a_s \exp{\left( 1 - \frac{t}{t_s} \right)^n}~.
\eea
From (\ref{asplit}) one sees that in a special case $n=0$, one recovers a standard big-bang scale factor with
\bea
\label{rhoBB}
\varrho_{BB}(t) &=& \frac{3}{8 \pi G(t)} \frac{m^2}{t^2}~~,\\
\label{pBB}
p_{BB}(t) &=& - \frac{c^2(t)}{8 \pi G(t)} \frac{m(3m-2)}{t^2}~,
\eea
and it is possible to write down an equation of state in the form of a barotropic perfect fluid as
\be
\label{bar}
p_{BB} = (\gamma - 1) \varrho c^2(t) = - \frac{c^2(t)}{8 \pi G(t) t^2} \frac{4(1-\gamma)}{3\gamma^2}~~,
\ee
where
\be
\label{gamdef}
\gamma = \frac{2}{3m}~~,
\ee
as in the standard notation \cite{PLB2011}. The standard big-bang models given by
(\ref{aBB}) are decelerating for $\gamma > 2/3$ $(m<1)$, and accelerating for $\gamma \leq 2/3$ $(m>1)$. The pressure is positive for $\gamma >1$ and negative for $\gamma < 1$. This of course also refers to its value at a big-bang singularity $(t=0)$, where it is plus infinity for $\gamma > 1$, and minus infinity for $\gamma < 1$.

Now, let us notice that in the limit $m=0$ the Eq.(\ref{newscalef}) reduces to an exotic singularity scale factor given by (\ref{aex}).
From (\ref{rhot}) and (\ref{pt}) we have
\bea
\label{exrhot}
\varrho_{ex}(t) &=& \frac{3}{8\pi G(t)} \frac{n^2}{t_s^2} \left( 1 - \frac{t}{t_s} \right)^{2(n-1)}~,\\
\label{expt}
p_{ex}(t) &=& - \frac{c^2(t)}{8 \pi G(t)} \times \\
&\times& \left[ 3 \frac{n^2}{t^2_s} \left( 1 - \frac{t}{t_s} \right)^{2(n-1)} + 2 \frac{n(n-1)}{t^2_s} \left( 1 - \frac{t}{t_s} \right)^{n-2} \right]~,\nonumber
\eea
and so
\bea
\label{exw}
w_{ex}(t) &=& \frac{p_{ex}(t)}{\varepsilon_{ex}(t)} = - \left[ 1 + \frac{2}{3} \frac{n-1}{n} \frac{1}{\left(1-\frac{t}{t_s}\right)^n} \right] \nonumber \\
&=& - \left[\frac{1}{3} - \frac{2}{3} q_{ex}(t) \right]~,
\eea
with the deceleration parameter equal to
\be
\label{exq}
q_{ex}(t) = - 1 - \frac{n-1}{n} \frac{1}{\left(1-\frac{t}{t_s}\right)^n}~~.
\ee
From (\ref{exrhot})-(\ref{exw}) we may conclude that for $n>2$ the energy density and pressure vanish $(\varrho=0$, $p=0)$, while the $w$-index blows-up to infinity which is exactly the characteristics of a $w$-singularity \cite{wsing}. In fact, the $w$-index blows-up for any positive value of $n>0$ $(n \neq 1)$, which together with the fact that the energy density diverges for $0<n<1$, and both the energy density and pressure diverge for $1<n<2$, means that we have neither FSF nor SFS singularity here.

Another way of writing $a_{BB}$ is the way one does in superstring cosmology \cite{PBB}
\be
a(t) = \Bigl\lvert \frac{t}{t_s} \Bigr\rvert^{\pm m}~~,
\ee
in which there are four branches (two of them ``pre-big-bang'').

\section{Regularizing singularities}
\setcounter{equation}{0}
\label{regular}

Despite that the definition of a singularity based on the Hawking and Penrose \cite{HE} geodesic incompletness has been a standard for years, today one is facing more subtleties in the matter. In fact, new types of singularities have been discovered (see e.g. Ref. \cite{aps10}) and what is more important, each of these singularities is characterized by different properties. First of all, lots of them do not even exhibit geodesic incompletness \cite{lazkoz} and so in view of the standard definition are not singularities at all! However, we realize that they allow a blow-up of various geometrical and physical quantities and may create a problem for the physical theories. For some of these singularities geodesics are not singular and still they have different "strength" which can be measured by various ingenious definitions. For example, a weak singularity definition of Tipler \cite{tipler+krolak} requires that the double integral
\be
\label{tipler}
\int_0^{\tau} d\tau' \int_0^{\tau'} d\tau'' R_{ab}u^a u^b~,
\ee
with $R_{ab}$ - the Ricci tensor, $u^a$ - the 4-velocity, and $\tau$ - the proper time, does not diverge on the approach to a singularity at $\tau = \tau_s$, while a weak singularity definition of Kr\'olak requires that the single integral
\be
\label{krolak}
\int_0^{\tau} d\tau' R_{ab}u^a u^b~~,
\ee
does not diverge on the approach to a singularity at $\tau = \tau_s$. Then, according to both of these definitions big-bang (type 0) and big-rip (type I) singularities are strong, while sudden future (type II), generalized sudden future (type IIg), big-separation (type IV), and $w$-singularities (type V) are weak. However, the finite scale factor singularities (type III) are strong with respect to Kr\'olak's definition (\ref{krolak}), and weak with respect to Tipler's definition. Bearing in mind another characteristic which is based on the spacetime averaging procedure proposed by Raychaudhuri \cite{raychaudhuri}, one learns that there are more subtleties \cite{PLB2011}. For example, a big-rip is a stronger singularity than a big-bang since its spacetime average blows-up, while for a big-bang it is not the case. On the other hand, sudden future singularities have vanishing spacetime average, while finite scale factor singularities might have an infinite spacetime average. Interestingly, in that sense the finite scale factor singularities can be considered stronger singularities than big-bang singularities. All this means that it is not obvious to say that replacing one singularity by another is like exchanging something physically troublesome into something else, which is also troublesome. There are subtleties, and so one should investigate the full nature of the physical object (a singularity) one deals with in order to learn its characteristics.

Motivated by the above discussion we present our main idea of the paper which is to investigate how different types of time-evolution of the physical constants like $c$ and $G$ influence the evolution of the universe. In particular, we are interested in a possibility to change the nature of singularities due to the variability of these constants. The review of the experimental bounds on the variability  of the constants are given in many references - the Ref. \cite{uzan} is one of the most recent ones.

It is quite reasonable that the nature of singularities may change while the constants are evolving. In fact, an analogous phenomenon, though due to a different physical reason - the conformal anomaly - results in strengthening a singularity \cite{houndjo}. Namely, an SFS singularity becomes an FSF singularity - the latter is a stronger type of singularity \cite{tipler+krolak,lazkoz,LFJ2007}. However, in our further investigations we will be discussing mostly weakening the singularities - this is a natural expectation in cosmology which helps to solve the singularity problem.

\begin{center}
{\bf a) a big-bang singularity}
\end{center}

For the purpose described above, we first suggest that the time-variation of the gravitational constant in (\ref{rhot}) and (\ref{pt}) of the form
\be
\label{GBB}
G(t) \propto \frac{1}{t^2}~~,
\ee
which is a faster decrease that in the standard Dirac's case $G(t) \propto 1/t$ \cite{dirac,alpha}, would presumably remove a big-bang singularity
in Friedmann models (i.e. removes both $p$ and $\varrho$ singularities). Such a time-dependence of $G$ would perhaps be less influenced by the geophysical constraints on the temperature of the Earth as it was discussed early in Ref. \cite{Teller}. On the other hand, in the Dirac's case $G(t) \propto 1/t$, only the $\varrho$ singularity can be removed.

Another suggestion is that the scale factor (\ref{newscalef}) would not approach zero at $t \to 0$ if it was rescaled be a ``regularizing'' factor $a_{rg} = (1+ 1/t^m)$ ($m \geq 0$), i.e.,
\be
a_{sm} = \left( 1 + \frac{1}{t^m} \right) \left(\frac{t}{t_s} \right)^m = \left(\frac{t}{t_s} \right)^m + \frac{1}{t^m_s}~~.
\ee

So far we have preliminary discussed a possibility for the gravitational constant to evolve in time. Now, we will also discuss the time-evolution of the speed of light. It is clear from (\ref{rhot})-(\ref{pt}) that any change of the type of a singularity which is also singularity of density cannot be done without admitting a curvature term in the Einstein equations (\ref{rho})-(\ref{p}). This, especially refers to a big-bang singularity - it cannot be removed at all, unless the spatial curvature is non-zero.

\begin{center}
{\bf b) other exotic singularities}
\end{center}

In order to regularize an SFS singularity by varying speed of light we suggest that the time-dependence of the speed of light is given by
\be
\label{regc}
c(t) = c_0 \left( 1 - \frac{t}{t_s} \right)^{\frac{p}{2}}~,
\ee
($c_0=$ const., $p=$ const.) which after substituting into (\ref{pt}) gives
\bea
\label{pregc}
&&p(t) = - \frac{c^2_0}{8\pi G} \left[ \frac{m(3m-2)}{t^2}\left( 1 - \frac{t}{t_s} \right)^p \right. \nonumber \\
&-& \left. 6 \frac{mn}{tt_s} \left( 1 - \frac{t}{t_s} \right)^{p+n-1}
+ 3 \frac{n^2}{t^2_s} \left( 1 - \frac{t}{t_s} \right)^{p + 2n -2} \right. \nonumber \\
&+&  \left. 2 \frac{n(n-1)}{t^2_s} \left( 1 - \frac{t}{t_s} \right)^{p+n-2} \right].
\eea
It then follows from (\ref{pregc}) that an SFS singularity is regularized by varying speed of light provided that
\be
p > 2 - n~~\hspace{0.5cm} (1<n<2)~~.
\ee
However, there is an interesting physical consequence of the functional dependence of the speed of light (\ref{regc}). Namely, it gradually diminishes reaching zero at the singularity. In other words, the light slows and eventually stops at an SFS singularity. Such an effect is predicted within the framework of loop quantum cosmology (LQC), where it is called the anti-newtonian limit $c = \sqrt{1-2 \varrho/\varrho_c} \to 0$ for $\varrho \to \varrho_c$ with $\varrho_c$ being the critical density \cite{mielczarek}. The low-energy limit $\varrho \ll \varrho_0$ gives the standard limit $c \to 1$.

One of the standard assumptions on the variation of the speed of light is that it follows the evolution of the scale factor \cite{VSL}
\be
\label{cc0}
c(t) = c_0 a^s(t)~~,
\ee
with $c_0$ and $s$ constant. The field equations (\ref{rho}) and (\ref{p}) become
\bea
\label{rhof}
\varrho(t) &=& \frac{3}{8\pi G(t)} \left(\frac{\dot{a}^2}{a^2} + kc_0^2 a^{2(s-1)} \right)~,\\
\label{pf} p(t) &=& - \frac{c_0^2 a^{2s}}{8\pi G(t)} \left(2 \frac{\ddot{a}}{a} + \frac{\dot{a}^2}{a^2} + kc_0^2a^{2(s-1)} \right)~.
\eea
With the time-dependence of $c(t)$ as in (\ref{cc0}), and with $a(t) = t^m$ it is possible to remove a pressure singularity provided $s > 1/m$ for $k=0$ and $m>0$, $s > 1/2$ or $m<0$, $s < 1/2$ for $k\neq 0$, but not the energy density singularity. A more sophisticated choice of the time variation of the speed of light which could regularize the singularities is
\be
\label{regchyb}
c(t) = c_0 \left( 1 - \frac{t}{t_s} \right)^{\frac{p}{2}} a^s(t)~,
\ee
or in terms of the scale factor
\be
\label{SFhyb}
a(t) = \left(\frac{c(t)}{c_0} \right)^{\frac{1}{s}} \left( 1 - \frac{t}{t_s} \right)^{-\frac{p}{2s}}~~.
\ee

Since $\varrho(t)$ does not depend on $c(t)$ (for $k=0$), then it is impossible to strengthen an SFS singularity to become an FSF singularity. It is possible only, if we assume that the gravitational constant $G$ changes in time. Let us then assume that
\be
\label{regG}
G(t) = G_0 \left( 1 - \frac{t}{t_s} \right)^{-r}~~,
\ee
($r=$ const., $G_0=$ const.) which changes (\ref{rhot}) and (\ref{pt}) to the form
\bea
\label{rhoG}
\varrho(t) &=& \frac{3}{8\pi G_0} \left[\frac{m^2}{t^2} \left( 1 - \frac{t}{t_s} \right)^r - \frac{2mn}{tt_s} \left( 1 - \frac{t}{t_s} \right)^{r+n-1} \right. \nonumber \\
&+& \left. \frac{n^2}{t^2_s} \left( 1 - \frac{t}{t_s} \right)^{r+2n-2} \right]~,\\
\label{pG}
p(t) &=& - \frac{c^2}{8\pi G_0} \left[ \frac{m(3m-2)}{t^2}\left( 1 - \frac{t}{t_s} \right)^r \right. \nonumber \\
&-& \left. 6 \frac{mn}{tt_s} \left( 1 - \frac{t}{t_s} \right)^{r+n-1} + 3 \frac{n^2}{t^2_s} \left( 1 - \frac{t}{t_s} \right)^{r + 2n -2} \right. \nonumber \\
&+& \left. 2 \frac{n(n-1)}{t^2_s} \left( 1 - \frac{t}{t_s} \right)^{r+n-2} \right]~.
\eea
From (\ref{rhoG}) and (\ref{pG}) it follows that an SFS singularity $(1<n<2)$ is regularized by varying gravitational constant when
\be
\label{con1}
r> 2-n~~,
\ee
and an FSF singularity $(0<1<n)$ is regularized when
\be
\label{con2}
r> 1-n~~.
\ee
On the other hand, assuming that we have an SFS singularity and that
\be
-1<r<0~,
\ee
we get that varying $G$ may change an SFS singularity onto a stronger FSF singularity when
\be
0 < r+n < 1~.
\ee
A physical consequence of the functional dependence of the gravitational constant in (\ref{regG}) is that the strength of gravity becomes infinite at the singularity. This is quite reasonable if we want to regularize an infinite (anti)tidal force at the singularity \cite{tipler+krolak}. This is also exactly what happens in the strong coupling limit $G \to \infty$ of gravity \cite{strongG}.

A hybrid case which would influence both types of singularities (big-bang (cf. (\ref{GBB})) and other exotic ones) is
\be
G(t) = \frac{G_0}{t^2} \left( 1 - \frac{t}{t_s} \right)^{-r}~,
\ee
which changes (\ref{rhot}) and (\ref{pt}) into
\bea
\label{rhoGh}
\varrho(t) &=& \frac{3}{8\pi G_0} \left[m^2 \left( 1 - \frac{t}{t_s} \right)^r - \frac{2mnt^2}{t_s} \left( 1 - \frac{t}{t_s} \right)^{r+n-1} \right. \nonumber \\
&+& \left. \frac{n^2t^2}{t^2_s} \left( 1 - \frac{t}{t_s} \right)^{r+2n-2} \right]~,\\
\label{pGh}
p(t) &=& - \frac{c^2}{8\pi G_0} \left[ m(3m-2)\left( 1 - \frac{t}{t_s} \right)^r \right. \nonumber \\
&-& \left. 6 \frac{mnt}{t_s} \left( 1 - \frac{t}{t_s} \right)^{r+n-1} + 3 \frac{n^2t^2}{t^2_s} \left( 1 - \frac{t}{t_s} \right)^{r + 2n -2} \right. \nonumber \\
&+& \left. 2 \frac{n(n-1)t^2}{t^2_s} \left( 1 - \frac{t}{t_s} \right)^{r+n-2} \right]~.
\eea
From (\ref{rhoGh}) and (\ref{pGh}) one can see that a big-bang singularity at $t=0$ is regularized with no additional conditions while SFS or FSF singularities are
regularized under the conditions (\ref{con1}) and (\ref{con2}) appropriately.

\begin{center}
{\bf c) singularities in (anti-)Chaplygin gas cosmology}
\end{center}

Since a couple of years, there has been a proposal that the dark energy can be simulated by a Chaplygin or an anti-Chaplygin gas model. One of the interests is that an anti-Chaplygin gas model allows the so-called big-brake singularity \cite{BBrake} ($\varrho \to 0$ and $p \to \infty$) which is a special case of a sudden future singularity $\varrho \to$ const. and $p \to \infty$. The equation of state of the (anti-)Chaplygin gas reads as
\be
p(t) = \mp \frac{A}{\varepsilon(t)} = \pm \frac{A}{\varrho(t)c^2(t)}~~,
\label{eoschap}
\ee
where $A >0$ is a constant with the unit of the energy density square (or pressure square) in $J^2 m^{-6} = kg^2 m^{-4} s^{-2}$, and the ``-'' sign refers to a Chaplygin gas, while the ``+'' sign refers to an anti-Chaplygin gas. It is interesting to note that the only limit of an SFS to get a big-brake for the scale factor (\ref{newscalef}) bearing in mind the fact that $\varrho \to 0$ (i.e., $\dot{a} \to 0$ - see (\ref{dotnew})) would be $m \to 0$ and $n \to 0$ which is not physically interesting. The reasonable limit of the scale factor (\ref{oldscalef}) is possible \cite{SFSobserv}.

Inserting (\ref{eoschap}) into (\ref{conser}) gives \cite{VSLChapGas}
\be
\label{conserCh}
\dot{\varrho}(t) + 3 \frac{\dot{a}}{a} \left(\frac{\varrho^2 c^4(t) \mp A}{\varrho(t) c^4(t)} \right) = - \varrho(t) \frac{\dot{G}(t)}{G(t)}
+ 3 \frac{kc(t)\dot{c}(t)}{4\pi G(t)a^2}~.
\ee
In order to find an exact solution, we will first assume that the speed of light is constant ($c_0=$ const.), and that the gravitational constant changes as
\be
\label{anG}
G(t) = G_0 \varrho(t)~,
\ee
with $G_0$ having the unit $m^3 kg^{-2} s^{-2}$, which gives (\ref{conserCh}) in the form
\be
\label{conserG}
\frac{2c_0^4 \varrho \dot{\varrho}}{\varrho^2 c_0^4 \mp A} = - 3 \frac{\dot{a}}{a}~~,
\ee
and integrates to give (we should make an assumption $A<\varrho^2c_0^4$ for ``-'' sign which corresponds to a Chaplygin gas)
\be
\label{rhosolnG}
\varrho(t) = \frac{\sqrt{\varrho_0}}{c_0^2} \left[ \pm A_c + a_0^3 \frac{1 \mp A_c}{a^3(t)} \right]^{\frac{1}{2}}
\ee
with $\varrho_0 =$ const. having the unit of the constant $A$, i.e., $J^2 m^{-6} = kg^2 m^{-4} s^{-2}$, $a_0$ (present value of the scale factor) having the unit of length, and $A_c \equiv A/ \varrho_0$. From (\ref{rhosolnG}) one can see that the standard big-bang limit $\varrho \sim a^{-3/2}$ is achieved if $A_c \to 0$ and $A_c \to \pm 1$ gives the cosmological constant limit. From (\ref{eoschap}) and (\ref{rhosolnG}) we have
\be
\label{psolnG}
p(t) = \mp c_0^2 \frac{\sqrt{\varrho_0}A_c}{\left[ \pm A_c + a_0^3 \frac{1 \mp A_c}{a^3(t)} \right]^{\frac{1}{2}}}~.
\ee
The standard big-bang limit $p \to 0$ is given for $A_c \to \pm 1$ and $p = \mp c_0^2 \sqrt{\varrho_0}$ for $A_c \to \pm 1$.

The more interesting solution in order to demonstrate regularization of singularities can be obtained in a general case of both varying $G=G(t)$ and $c=c(t)$ though with zero curvature $k=0$ case by making the following ansatz in (\ref{conserCh})
\be
\label{ansatzc}
\varrho(t) c^2(t) = B = {\rm const.}~~,
\ee
with $B$ having the unit of energy density, which gives
\be
\label{conserc}
\frac{\dot{\varrho}}{\varrho} + 3 \frac{\dot{a}}{a} \left( \frac{B^2 \mp A}{B^2} \right) + \frac{\dot{G}}{G} = 0~.
\ee
The solution of (\ref{conserc}) reads as
\be
\label{solconsc}
\varrho(t) a^{3\gamma}(t) G(t) = E = {\rm const.}~,
\ee
where we have defined
\be
\label{defgam}
\gamma \equiv \frac{B^2 \mp A}{B^2}
\ee
for the sake of comparison with the standard cosmology (here $\gamma$ has not the meaning of the barotropic index, but as we can learn it makes the scale factor scaling the same as in standard case). Bearing in mind (\ref{ansatzc}) and (\ref{solconsc}) we obtain that
\be
\label{c}
c^2(t) = \frac{B}{E} a^{3\gamma}(t) G(t)~~.
\ee
From (\ref{eoschap}) and (\ref{c}) we finally have
\bea
\label{finalp}
p(t) &=& \mp \frac{A}{B} = {\rm const.} ~~,\\
\label{finalrho}
\varrho(t) &=& \frac{B}{c^2(t)} = \frac{E}{a^{3\gamma}(t) G(t)}~~.
\eea
Putting the standard big-bang scale factor (\ref{aBB})
\be
a(t) = \left( \frac{t}{t_s} \right)^m = \left( \frac{t}{t_s} \right)^{\frac{2}{3\gamma}}~~,
\ee
we now have from (\ref{finalp}) and (\ref{finalrho})
\bea
\varrho(t) &=& \frac{Et_s^2}{t^2G(t)}~~,\\
p(t) &=& \mp \frac{A}{B} = {\rm const.} ~~,
\eea
which give $\varrho \to \infty$ and $p(0) = 0$, provided $G(0) =$ const. $\neq 0$. The singularity at $t=0$ in $\varrho$ can be regularized by taking $G(t) \propto 1/t^2$ as in (\ref{GBB}). In our case we have a constant pressure (cosmological term) instead of zero pressure.

\section{Conclusions}
\setcounter{equation}{0}
\label{conclusion}

We have shown by specifying some examples that it is possible to regularize cosmological singularities due to variation of the physical constants. Although it seems to work in a general framework of varying constants theories, we have considered this phenomenon in the theories with varying speed of light (VSL) $c(t)$, and with varying gravitational constant $G=G(t)$.

For example, in order to regularize a big-bang singularity, the simple modification of Dirac's relation ($G \propto 1/t^2$ instead of $G \propto 1/t$) is required. In order to regularize a sudden future singularity (SFS), finite scale factor singularity (FSF) or a $w-$singularity, some more complicated time-dependence for $c(t)$ and $G(t)$ is necessary. We have found such a dependence and it was appropriate to a newly introduced scale factor given by (\ref{newscalef}). Interestingly, in order to regularize an SFS by varying $c(t)$, the light should stop propagating at a singularity - the fact which appears in the loop quantum cosmology \cite{mielczarek}. On the other hand, to regularize an SFS by varying gravitational constant - the strength of gravity has to become infinite at a singularity (as in the strong coupling limit of gravity \cite{strongG}) which is quite reasonable because of the requirement to overcome the infinite (anti-)tidal forces at singularity.

We have also studied the regularization of singularities by varying $c(t)$ and $G(t)$ within the framework of
(anti-) Chaplygin gas cosmology and have shown that regularization is also possible in these theories. 

In view of the variation of the velocity of light $c(t)$ there is a crucial difference between the mass density $\varrho$ and the energy density $\varepsilon = \varrho c^2$, since variation of $c(t)$ effects the Einstein mass energy formula $E=mc^2$ transformed here as the
the mass density versus pressure formula $p = \varrho c^2$ after dividing both sides by the volume. Then, if one takes into account a barotropic equation of state, then
it is better to define the barotropic index which is dimensionless, as we did in (\ref{wute}). Since the pressure has the same units as the energy density, and the energy density results in multiplying the mass density by $c^2(t)$, then it is more reasonable to talk about singularities in the mass density and the pressure rather than in the energy density and pressure since they are, in fact, equivalent. The only factor which relates them is the barotropic index which we have assumed to be dimensionless. In other words, the power to remove a singularity by varying speed of light $c=c(t)$ refers only to the pressure, and not to the mass density. This can be seen from Eqs. (II.11)-(II.12). Taking into account any explicit form of matter such as for example the radiation $p(t) = (1/3) \varrho c^2(t)$ one easily sees that regularization with $c^2(t)$ can be done for the pressure only. The mass density cannot be regularized this way.

In our paper we do not claim that we have solved fully the problem of singularity due to variation of the physical constants. Rather we first give an idea or a path for other cosmologists to follow. There is a subtlety in our approach, since we replace singularities in geometrical quantities by kind of singularities in physical fields (both constants $c(t) \to 0$ and $G(t) \to \infty$ can be treated as such), but it is not yet known whether these new singularities are more harmful than the original ones. For example, in the low-energy-effective superstring theory, there are two kind of singularities: the curvature singularities and the strong coupling singularities for which the dilaton may diverge. They are not necessarily of the same type though both can be regularized by the quantum corrections. Anyway, the singularities we have presented and regularized in our paper are characterized by many different properties. Lots of them do not even exhibit geodesic incompletness and so in view of the standard definition they are not singularities at all. However, they still have some other characteristics which are divergent and, what is more interesting, they have different "strength" which means they are less or more harmful and our ``regularization'' approach may lead to a change of this strength.

Finally, we hope that variation of the physical constants which leads to regularization of singularities may be useful in the discussions of the multiverse concept giving the link through a kind of ``fake'' singularities to various parts of the universe with different physics. Of course our discussion is preliminary and should be continued by using appropriate mathematical formalism of both general relativity and particle physics.

\section{Acknowledgements}

The authors acknowledge the discussions with Adam Balcerzak, Tomasz Denkiewicz, and Jakub Mielczarek. We also thank Alexander Kamenshchik for informing us
about Chaplygin gas equation of state in relation to Ref. \cite{BBrake}. This work was supported by the National Science Center grant No N N202 3269 40 (years 2011-2013).

\end{document}